\documentclass[twocolumn]{article}

\usepackage[T1]{fontenc}
\usepackage{geometry}
\usepackage{lipsum}
\usepackage{graphicx}
\usepackage{caption}
\usepackage[dvipsnames,svgnames,table]{xcolor} 

\usepackage{hyperref}
\hypersetup{
  linkcolor  = ForestGreen,
  citecolor  = BlueGreen,
  urlcolor   = RedOrange,
  colorlinks = true,
}

\usepackage{amssymb}  
\usepackage{amsthm}
\usepackage{amsmath}
\usepackage{float}
\usepackage{siunitx}
\usepackage{authblk} 

\usepackage{xspace}
\newcommand\tspect{$\tau$SPECT\xspace}
\newcommand\pentrack{{\ttfamily PENTrack}\xspace}
\newcommand\mcucn{{\ttfamily MCUCN}\xspace}
\newcommand\magpylib{{\ttfamily magpylib}\xspace}
\newcommand\penconf{{\ttfamily penconf}\xspace}
\newcommand\penplot{{\ttfamily penplot}\xspace}

\usepackage[T1]{fontenc}

\usepackage[%
  hyperref=true,
  backend=biber,
  url=true,
  isbn=false,
  backref=false,
  citestyle=numeric-comp,
  doi=true,
  uniquelist=false,
  maxcitenames=1,
  maxbibnames=5,
  sorting=none,
  block=none
]{biblatex}

\AtEveryBibitem{%
  \clearfield{title}
}

\AtEveryBibitem{
    \clearfield{urlyear}
    \clearfield{urlmonth}
}

\renewbibmacro{journal}{%
  \ifboolexpr{
    test {\iffieldundef{shortjournal}}
    and
    test {\ifentrytype{article}}
  }
  {\printfield{journaltitle}}
  {\printfield{shortjournal}}
}

\title{Ultra-cold neutron simulation framework for the free neutron lifetime experiment \tspect}

\author[1]{J.~Auler}
\author[1]{U.~Bajpai}
\author[2]{M.~Engler}
\author[1]{V.~Ermuth}
\author[1,b]{M.~Fertl}
\author[2]{K.~Franz}
\author[1]{W.~Heil}
\author[2]{S.~Kaufmann}
\author[3]{B.~Lauss}
\author[1]{N.~Pfeifer}
\author[2,3,c]{D.~Ries}
\author[1,a]{S.~Vanneste}
\author[2,3]{N.~Yazdandoost}

\affil[ ]{The \tspect collaboration}
\affil[ ]{ }

\affil[1]{\small Institute of Physics, Johannes Gutenberg University Mainz, 55099 Mainz, Germany}
\affil[2]{\small Department of Chemistry - TRIGA site, Johannes Gutenberg University Mainz, 55099 Mainz, Germany}
\affil[3]{\small Paul Scherrer Institut, CH-5232 Villigen PSI, Switzerland}

\affil[ ]{ }

\affil[a]{Corresponding author: \texttt{svannest@uni-mainz.de}}
\affil[b]{Corresponding author: \texttt{mfertl@uni-mainz.de}}
\affil[c]{Corresponding author: \texttt{dieter.ries@psi.ch}}

\begin{document}

\onecolumn

\maketitle

\begin{abstract}
  

The precise determination of the free neutron lifetime is of great significance in modern precision physics. This key observable is linked to the mixing of up and down quarks via the Cabibbo-Kobayashi-Maskawa matrix element  $V_{ud}$, and the abundance of primordial elements after the Big-Bang Nucleosynthesis. However, the two leading measurement techniques for the neutron lifetime currently yield incompatible results, a discrepancy referred to as the neutron lifetime puzzle. To address the systematic uncertainties arising from neutron interactions with material walls, the \tspect experiment employs a fully magnetic trap for ultra-cold neutrons (UCNs).

UCNs are extremely low-energy neutrons with typical velocities below $8\,\textrm{m/s}$, which can be manipulated using magnetic fields, gravity, and suitable material guides, whose surface can reflect them at any angle of incidence. To precisely study and characterize UCN behavior during production, guidance, storage, and detection in \tspect, we have developed a dedicated simulation framework. This framework is built upon the externally developed UCN Monte Carlo software package \pentrack and is enhanced with two companion tools: one for flexible and parametrizable upstream configuration of \pentrack such that the simulation's input settings can be adjusted to reproduce the experimental observations. The second package is used for analyzing, visualizing, and animating simulation data.

The simulation results align well with experimental data obtained with \tspect at the Paul Scherrer Institute and serve as a powerful resource for identifying systematic uncertainties and guiding future improvements to the current experimental setup.

\end{abstract}

\twocolumn

\section{Introduction and motivation}
\label{sec:introduction}

The free neutron lifetime, $\tau_{n}$, is fundamentally connected to fundamental parameters of the Standard Model (SM) of particle physics and cosmological observables. Regarding the SM, a precise measurement of $\tau_{n}$, in combination with the ratio of axial-vector to vector coupling $\lambda\equiv g_{A}/g_{V} $, can provide a determination of the first element of the Cabibbo–Kobayashi–Maskawa (CKM) quark mixing matrix, $V_{ud}$, without the nuclear structure corrections needed for the interpretation of $0^{+}\rightarrow 0^{+}$ super-allowed decays~\cite{hardySuperallowed002020,  sengUpdateSemileptonicKaon2022, doe/nsfnuclearscienceadvisorycommittee&longrangeplanworkinggroupNewEraDiscovery2023}. In cosmology, the lifetime $\tau_{n}$ directly influences the number of available neutrons during Big-Bang Nucleosynthesis. A precise measurement of $\tau_{n}$ is required to estimate the primordial nuclei abundances in the early Universe, which sets the initial conditions for large scale structures formation and evolution, i.e.~\cite{mathewsBigBangNucleosynthesis2005, chowdhuryNeutronLifetimeAnomaly2024}.

Most measurements of $\tau_{n}$ are achieved using either of two distinct techniques. The beam method relies on counting the number of $\beta$-decaying neutrons through the process $n\to p^{+} + e^{-} + \bar \nu_{e}$. This is typically achieved by counting the number of charged particles, electrons $e^{-}$, or protons $p^{+}$, exiting from a predetermined volume through which a neutron beam is passing. Those counts are then compared to the neutron flux of the beam for determining $\tau_{n}$.

The bottle method relies on counting and comparing the number of remaining neutrons after trapping and storing them in a volume for different time periods. Those measurements rely on ultra-cold neutrons (UCNs), whose velocities are below \SI{8}{\meter/\second}, which can be manipulated using magnetic fields, gravity, and suitable material guides, whose surface can reflect them at any angle of incidence.

As of today, both experimental techniques provide incompatible results, giving rise to the so-called free neutron lifetime puzzle~\cite{paulPuzzleNeutronLifetime2009, wietfeldtColloquiumNeutronLifetime2011, greeneNeutronEnigma2016, ezhovMeasurementNeutronLifetime2018, serebrovNewNeutronLifetime2018, broussardNewSearchMirror2019, berezhianiNeutronLifetimePuzzle2019, hirotaNeutronLifetimeMeasurement2020, serebrovSearchExplanationNeutron2021, wietfeldtCommentsSystematicEffects2022, wietfeldtNeutronLifetimeDiscrepancy2024, kochExcitingHintSolution2024}. The current most precise determination of $\tau_{n}$ is given by the magneto-gravitational trap, UCN$\tau$, with $\tau_{\textrm{n,UCN}\tau} = 877.82\,\textrm{s} \pm 0.22_{-0.17}^{+0.20}\,\textrm{s}$~\cite{pattieMeasurementNeutronLifetime2018, gonzalezImprovedNeutronLifetime2021, musedinovicMeasurementFreeNeutron2024}. Compared to material bottles, systematic uncertainties due to UCN up-scattering or absorption on material walls during storage in magnetic traps are  avoided, leading to a more comprehensive and precise measurement of $\tau_{n}$. The experiment \tspect, whose simulation framework is presented in this document, relies on a fully magnetic bottle experiment to confine neutrons and to measure $\tau_{n}$~\cite{aulerSPECTSpinflipLoaded2024}. \tspect aims to provide an independent cross-check measurement of $\tau_{n}$, with a sensitivity reach of $\sigma_{\tau_{n}} \leq 0.3 \, \textrm s$ in the near future. Reaching this goal requires both an overall high number of trapped UCNs, and precise identifications of sources of systematic uncertainties. With this ambitious goal in mind, an end-to-end simulation framework encapsulating the process of UCN production, guiding, trapping, and detection inside \tspect has been developed, and is being presented here.

In Sec.~\ref{sec:simframework}, a description of the \tspect experiment and the integration of its various subsystems in the simulation framework is presented. A selection of end-to-end simulation results are then presented in Sec.~\ref{sec:endtoendsimulation}. We conclude in Sec.~\ref{sec:conclusion} with a summary and a discussion of potential future applications of the simulation framework.

\begin{figure*}[pht]
  \centering
  \includegraphics[width=1\textwidth, trim={1.5cm 3.5cm 3cm 7cm},clip]{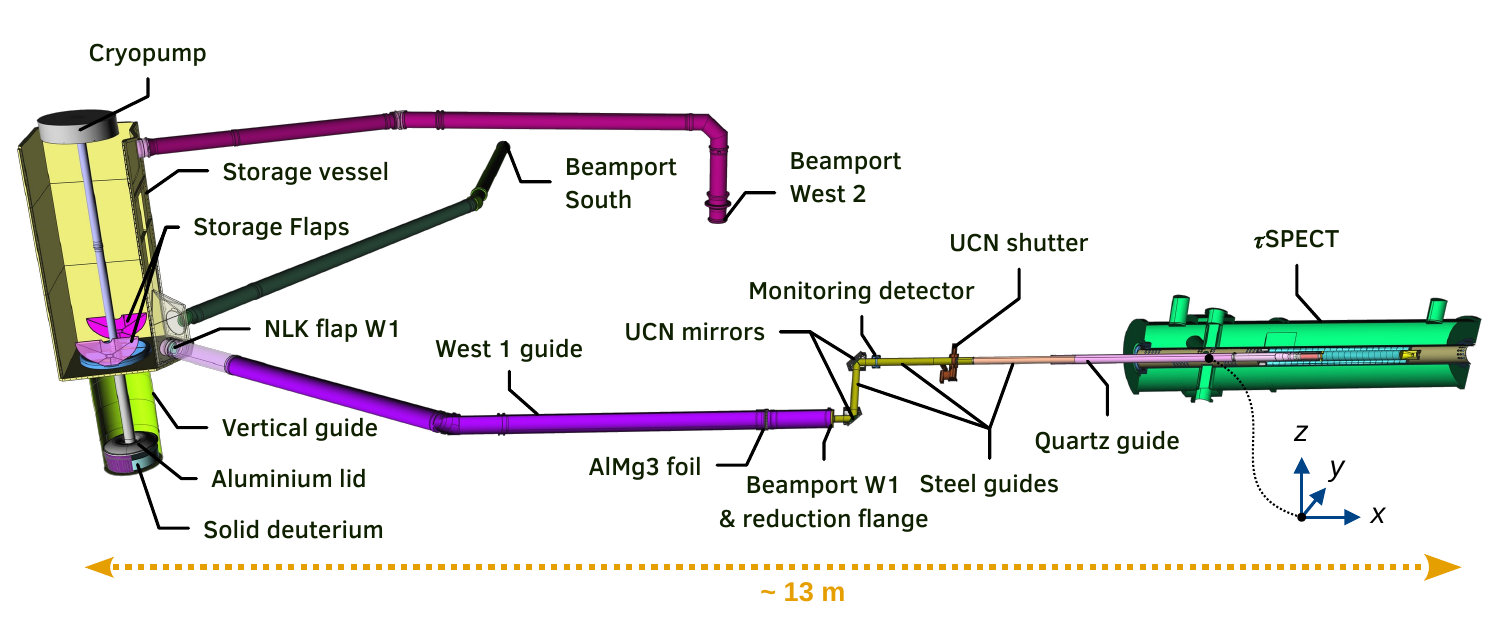}
  \caption{Cut-view of the \tspect experiment attached to the UCN source at PSI. In the simulations, neutrons originate from the solid deuterium, pass through an aluminum lid, and are guided vertically up before entering the storage vessel, which is then closed by mechanical flappers. UCNs are delivered using a horizontal guide to the beamport West 1 (W1), to which a reduced diameter flange is connected. UCNs are guided upward via a stainless steel beamline featuring two UCN mirrors, before passing a UCN shutter and finally reaching \tspect.}
  \label{fig:PSI_tspect}
\end{figure*}

\begin{figure*}[pht]
  \centering
  \includegraphics[width=1\textwidth, trim={0 2.8cm 0 3.5cm},clip]{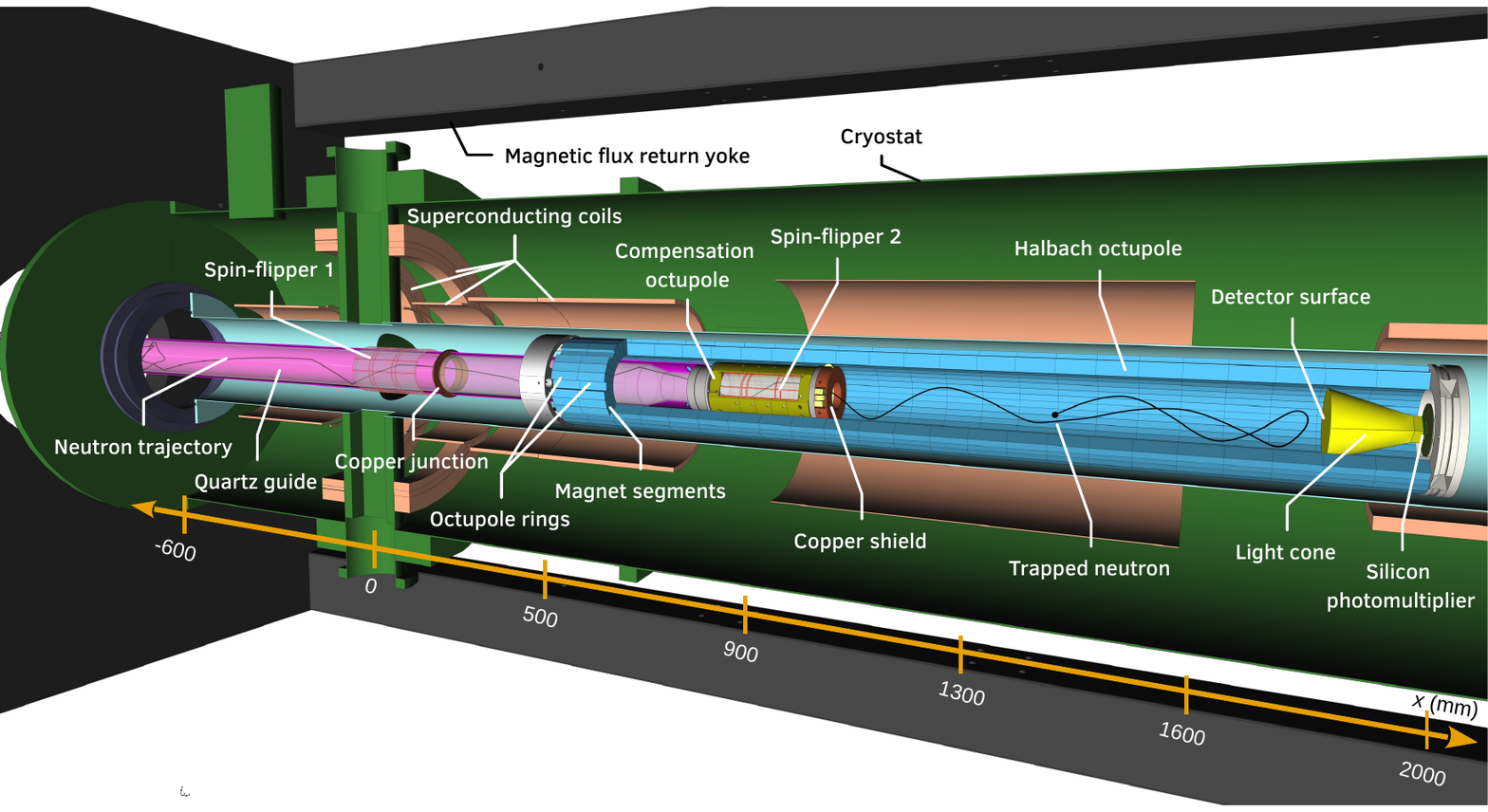}
  \caption{Cut view representation of \tspect. The axial magnetic field is generated by super conducting coils in a cryostat. The transverse magnetic field is generated by a permanent Halbach octupole magnet. UCNs enter the experiment from the left, and are guided in quartz tubes. Their polarization is adiabatically reversed, bringing them into a storable state, using spin-flipper units installed around the guides which can be moved in and out of the storage volume. A second octupole is installed around the spin-flipper 2 for compensating the main octupole and assure better spin-flip efficiency. A neutron detector is mounted on a translation stage that can enter the storage volume from the right to absorb and count neutrons in the trap.}
  \label{fig:tspectExperiment}
\end{figure*}

\section{Framework components}
\label{sec:simframework}

The framework was implemented using the C++ UCN Monte Carlo simulation software \pentrack~\cite{schreyerPENTrackSimulationTool2017}. It simulates dynamics of UCNs and their decay products in complex geometries and electromagnetic fields. A wide number of parameters and properties can be tracked over time, to mention a few: particle position, speed, kinetic and potential energies, spin orientation, surface hits.

\pentrack is extensively customized internally to fit \tspect requirements. Among others, we implemented the following options: multi-threading; virtual spin-flip surfaces that reverse the polarization of passing neutrons; adiabatic fast passage flipping of the neutron spin; sampling an external source file of initial parameters for the UCNs; loading of multiple wire segments accepting alternating and direct currents for generating magnetic fields.

Moreover, two python companion software packages have been developed and used to produce the results presented in this document, discussed in Sec.~\ref{sec:endtoendsimulation}: \penconf\footnote{\href{https://gitlab.rlp.net/tauSPECT/penconf}{https://gitlab.rlp.net/tauSPECT/penconf}}, from which the \tspect experimental settings were virtually reconstructed, and which offers flexible upstream generation of \pentrack configuration files; and \penplot\footnote{\href{https://gitlab.rlp.net/tauSPECT/penplot}{https://gitlab.rlp.net/tauSPECT/penplot}}, working downstream, and focusing on simulation output handling, filtering, analysis, geometry visualization, 3D plots and animations.

Magnetic field meshes are built externally, using the \magpylib python package~\cite{ortnerMagpylibFreePython2020}, further discussed in Sec.~\ref{sec:magneticfield}.

Material properties for each surface and volume are assigned in a configuration file loaded in \pentrack, see Appendix~\ref{sec:materials}.

The framework includes the Paul Scherrer Institute (PSI) UCN source~\cite{ bisonNeutronOpticsPSI2020, bisonUltracoldNeutronStorage2022}. \tspect is connected to the beamport West 1 (W1) since summer 2023, as shown in Fig.~\ref{fig:PSI_tspect}, and detailed in Sec.~\ref{sec:psisource}.

UCNs are guided from the PSI source to \tspect using a \emph{mirror} beamline (discussed in Sec.~\ref{sec:beamline}, shown in Fig.~\ref{fig:CADmonitordet}) using stainless steel and quartz tubes. A monitoring detector in the beamline counts a small subset of neutrons for each run. The main components of the \tspect apparatus are shown in Fig.~\ref{fig:tspectExperiment}. Two spin-flipping (SF) units consist of coils fed by radio-frequency (RF) signals, discussed in Sec.~\ref{sec:sf}. They manipulate the neutron spin along the neutron path, converting UCNs into a storable state relative to the magnetic trap potential formed by superconducting coils and a permanent Halbach octupole magnet, as described in Sec.~\ref{sec:magneticfield}. After a filling period, the SF units attached to a translation stage are retracted from the trap volume, leaving UCNs in a material-free environment. After a defined storage time, the remaining neutrons are counted (and absorbed) using an in-situ detector that can be translated in and out of the trap. Measurement runs are then repeated for different storage intervals. The resulting (exponential) decay curve over the storage time can then be fitted to extract the free neutron lifetime.

\subsection{PSI UCN source}
\label{sec:psisource}

Over the years, the UCN group at PSI has developed a detailed Monte-Carlo simulation framework, \mcucn~\cite{zsigmondMCUCNSimulationCode2018}, that has been optimized to reflect the operation behavior of the UCN source~\cite{bisonNeutronOpticsPSI2020, bisonUltracoldNeutronStorage2022}


Under nominal conditions, the $2.2\,\textrm{mA}$ $590\,\textrm{MeV}$ proton beam from PSI's High Intensity Proton Accelerator (HIPA)~\cite{grillenbergerHighIntensityProton2021} is directed towards a neutron spallation target for \SI{8}{\second}. The neutrons are first moderated in heavy water before they are down-converted to UCNs in solid deuterium ($\textrm{sD}_{2}$). After leaving the $\textrm{sD}_{2}$, UCNs traverse an $\textrm{AlMg}_{3}$ lid and are guided vertically up into a storage vessel. After those $8\,\textrm s$, flappers are closed, preventing UCNs to leave the storage vessel toward the bottom. Three UCN guides lead from the storage vessel to three different beamports and UCN experiments.

We integrated the PSI UCN source into our simulation framework to accurately reproduce its well-established properties, as documented in \cite{atchisonTransmissionVerySlow2009, bisonNeutronOpticsPSI2020, bisonUltracoldNeutronStorage2022, bisonCharacterizationUltracoldNeutron2023}. The simulation starts with UCNs exiting the $\textrm{sD}_{2}$ surface, during a period of $8\,\textrm s$.
Experimentally, the distribution of the initial neutron kinetic energy, $E_k$, has been shown to follow a power law $E_k^{2.7}$~\cite{bisonNeutronOpticsPSI2020}. When exiting the $\textrm{sD}_{2}$ bulk material, each UCN gets an additional kinetic energy boost of about $105\,\textrm{neV}$ as its potential energy is converted into kinetic energy~\cite{altarevDirectExperimentalVerification2008}. UCNs re-entering the $\textrm{sD}_{2}$ volume are considered to be lost.

\begin{figure}[t]
  \centering
  \includegraphics[width=0.49\textwidth, trim={1.0cm 1.6cm 1cm 3.5cm},clip]{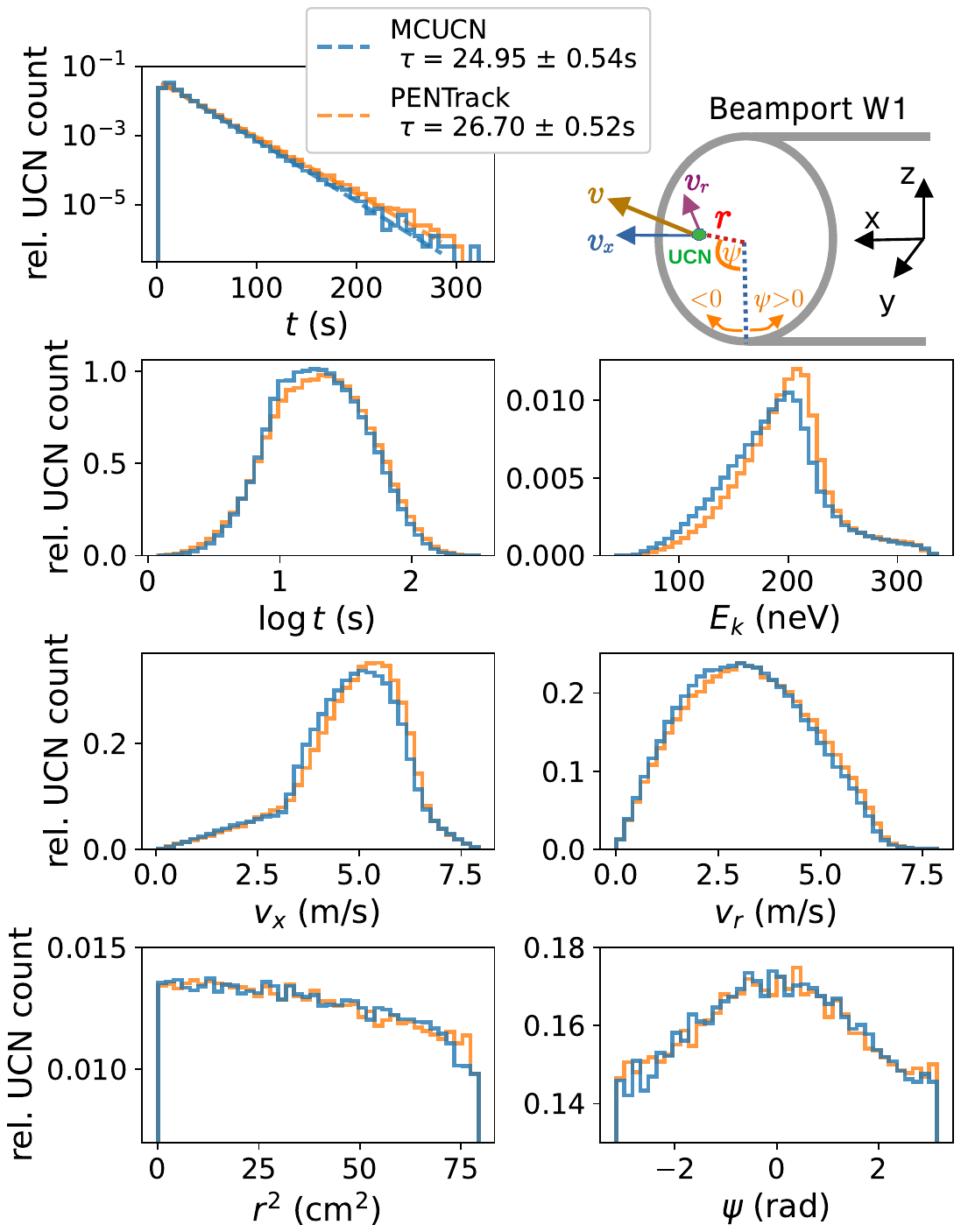}
  \captionsetup{width=.9\linewidth}
  \caption{Comparison between \mcucn$^{\ref{fn:refGeza}}$ and \tspect simulation frameworks (using \pentrack) of the neutron distributions at beamport West\,1, for a UCN production cycle (starting at $t=0\, \textrm s$) in the PSI source. UCNs reaching the exit are recorded and terminated. Their positions and velocities are shown in polar coordinates ($r$-$\psi$). The velocity $v$ is decomposed into a longitudinal component $v_{x}$ and perpendicular component $v_{r}$. $E_k$ is the kinetic energy. The storage time constant, $\tau_{s}$, is defined in Eq.~\eqref{eq:tauStorage}.}
  \label{fig:BeamportW1mcucnVSpentrack}
\end{figure}

A comparison of the results\footnote{\label{fn:refGeza}\mcucn results provided by Geza Zsigmond, private communication.} of the UCN source from both simulation frameworks has been performed. Fig.~\ref{fig:BeamportW1mcucnVSpentrack} shows characteristic properties of the UCNs arriving at the beamport West 1. Both frameworks provide similar distributions of UCN arrival positions, time, velocities, and energy. After the source storage vessel is filled, the arrival time distribution of UCNs at the beamport, $N(t)$, and the storage time constant, $\tau_{s}$, can be described by an exponential law,
\begin{equation}\label{eq:tauStorage}
  N(t) = N_{0}e^{-t/\tau_{s}},
\end{equation}
with $N_0$ the initial number of UCNs at the time $t=0$. The minute differences in the distributions and the storage time constants, $\tau_{\textrm{MCUCN}} = 24.95\,\textrm{s} \pm 0.54\,\textrm{s}$ and $\tau_{\textrm{PENTrack}} = 26.70\,\textrm{s} \pm 0.52\,\textrm{s}$, could be explained by the slightly different geometries, small gap locations, and material properties.

\subsubsection{Virtual source}

Starting UCNs from the $\textrm{sD}_{2}$ crystal for each simulation run is computationally cumbersome and expensive, as only a very small fraction of them ultimately reach \tspect. Moreover, we have verified in the simulations that modifications of \tspect's configuration, e.g., its height relative to the beamport, barely influences the dynamics of the neutrons in the source. We can therefore separate both volumes, the UCN source and \tspect, and consider them as independent. This separation is applied after the beamport and the diameter reduction flange, see Fig.~\ref{fig:CADmonitordet}.

Simulations of UCN dynamics can thus be started before the bottom mirror of the beamline connecting \tspect with the beamport. To do so, the computation is split into two steps:

\begin{enumerate}

\item Virtual detector recording: an initial global simulation is computed, which features the source, the beamline, and \tspect. A virtual detector plane placed at the separation between the source and the beamline records the relevant properties of all passing neutrons without interfering with their dynamics.

\item Virtual source sampling: the resulting UCN distribution at the virtual detector can then be sampled for simulations focusing on \tspect only. In that case, the PSI source volume is removed. We verified with the global simulation that most UCNs that get reflected to the PSI source do not go back into \tspect for a second time. The few neutrons that do get back into \tspect are nevertheless counted for each new pass during the step 1 (virtual detector recording), and their statistics are therefore also accounted for during the virtual source sampling phase. Consequently, we can add a neutron absorbing plane right before the virtual detector. This ensures that neutrons that get reflected from \tspect back to the PSI source are terminated. 

\end{enumerate}

UCNs generally undergo some amount of diffuse scattering on surfaces, which is implemented into \pentrack as a random process. After a few reflections, neutrons starting with the same initial conditions will end up in completely different positions in the UCN phase space volume. The virtual source sampling technique introduced above allows for the same initial conditions to be sampled several times (within reasonable limits) during a simulation, while still leading to a variety of different final neutron states.

\subsection{Beamline and monitoring detector}
\label{sec:beamline}
\begin{figure}[th]
  \centering
  \includegraphics[width=0.49\textwidth, trim={1cm 1cm 1cm 1cm},clip]{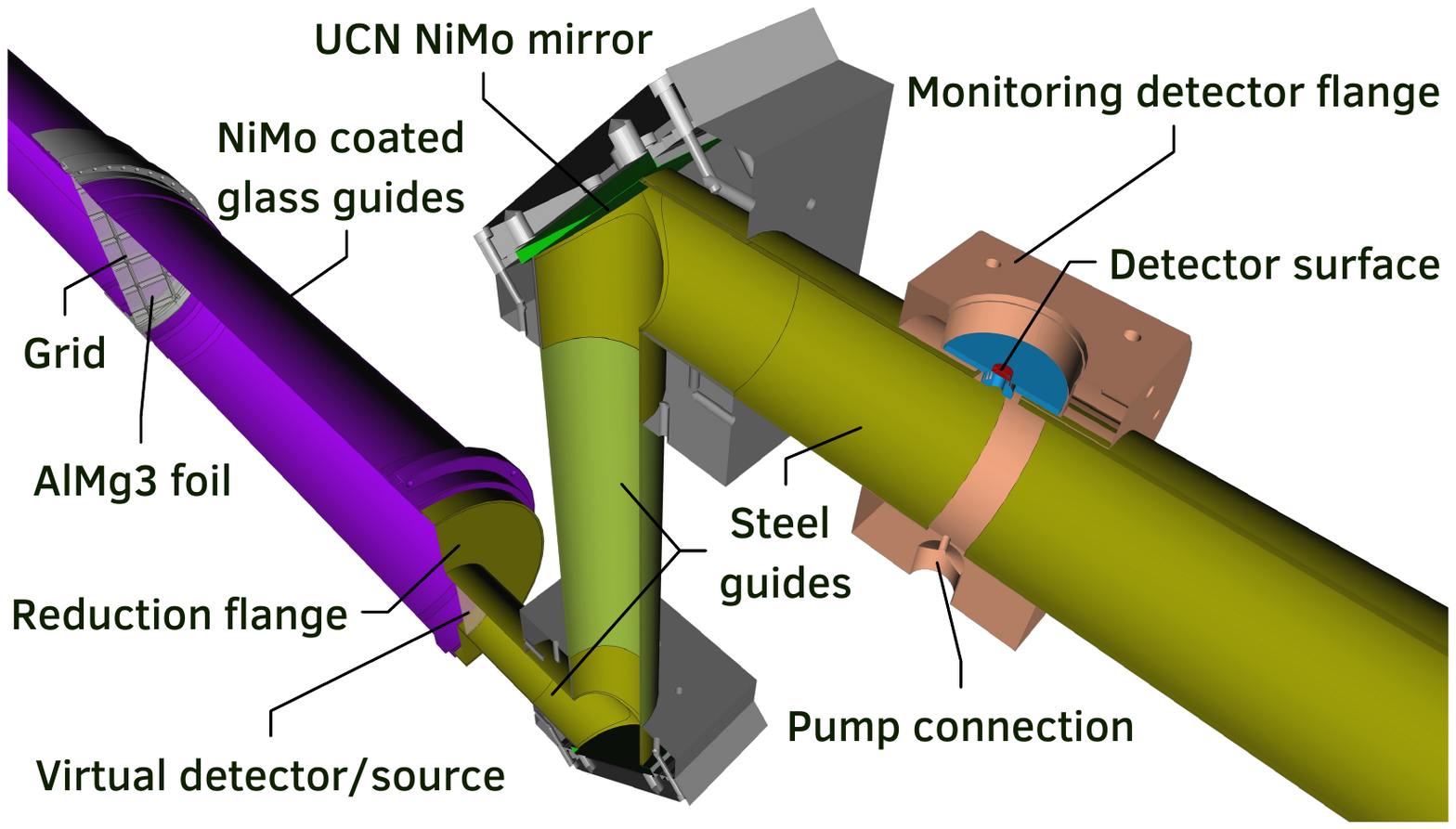}
  \captionsetup{width=.9\linewidth}
  \caption{CAD cut view of beamport W1, the mirror beamline, and the monitoring detector. The virtual detector plane records every neutron pass, which are later sampled for starting UCN simulations directly from the beamport.}
  \label{fig:CADmonitordet}
\end{figure}

The \tspect apparatus is mounted on four vertical lifters. By varying the height of \tspect relative to the beamport exit, thus slowing down UCNs gravitationally (about $102\, \textrm{neV/m}$), different parts of the initial neutron energy spectrum, shown in Fig.~\ref{fig:BeamportW1mcucnVSpentrack}, can be loaded. The beamline, displayed in Fig.~\ref{fig:CADmonitordet}, consists of vertical and horizontal stainless steel tubes with an inner diameter of $66\,\textrm{mm}$ connecting beamport W1 to the entrance of \tspect. Both 90° turns comprise \emph{mirror} plates $^{58}\textrm{NiMo (85:15 weight ratio)}$ coated, which assure a high UCN reflectivity. The simulation framework can account for any beamline \emph{height} (defined as centerline offset between top and bottom horizontal guides), ranging from $0\,\textrm{m}$ up to $2\,\textrm{m}$ in practice. New designs of beamline geometries and materials can be tested and optimized as well.

A monitoring detector is installed after the top mirror and before the UCN shutter. It tracks fluctuations in the total number of UCNs produced for each individual pulse of the PSI source. The number of neutrons counted after storage inside \tspect can then be normalized based on the monitoring detector counts. The monitoring detector consists of a stainless steel flange with a $4\,\textrm{mm}$ diameter pinhole that allows a small fraction of the UCNs to reach a detector surface.

\begin{figure}[th]
  \raggedleft
  \includegraphics[width=0.49\textwidth, trim={0cm 0cm 0cm 0cm},clip]{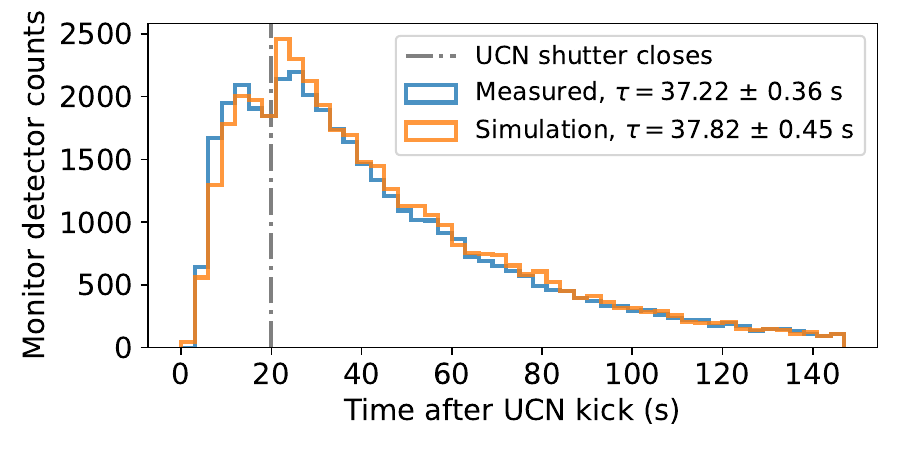}
  \captionsetup{width=.9\linewidth}
  \caption{Comparison between simulation and measured time distributions of UCNs counted by the monitoring detector, with a bin size of $3\,\textrm s$. In this example, two measurement cycles are stacked to increase statistics. \tspect is at a height of $43\,\textrm{cm}$. Its UCN shutter closes after $20\,\textrm s$, after which neutrons stay confined in the beamline which is still exchanging neutrons with the PSI UCN source. The corresponding storage time constant $\tau$ is extracted by fitting an exponential function starting at \SI{25}{\second} as Eq.~\eqref{eq:tauStorage}.}
  \label{fig:timdeDistMonDet}
\end{figure}

The monitoring detector counting distributions over time are compared between simulation and experimental measurement in Fig.~\ref{fig:timdeDistMonDet}. Both distributions are in excellent agreement, and show similar UCN storage time constants $\tau$ of the beamline after the UCN shutter in front of \tspect (see Fig.~\ref{fig:PSI_tspect}) is closed.

\subsection{Magnetic trap}

\label{sec:magneticfield}

\begin{figure}[ht]
  \centering
  \includegraphics[width=0.49\textwidth, trim={0 0.2cm 0 0cm},clip]{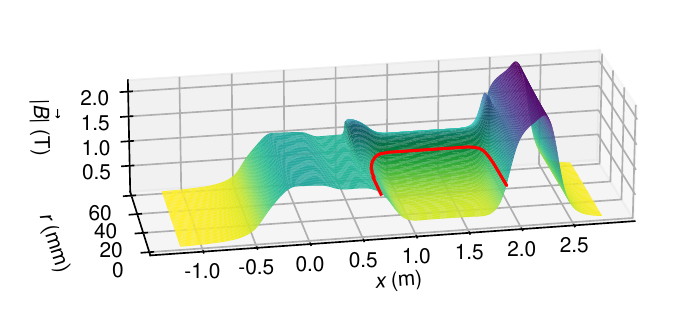}
  \captionsetup{width=.9\linewidth}
  \caption{Magnetic field amplitude $|\vec B|$, along the longitudinal direction $x$, and radial direction $r$. The longitudinal field is produced by superconducting coils, while most of the contribution to the transversal field comes from the permanent Halbach octupole magnet. The edge of the UCN trapping region is indicated by a red line.}
  \label{fig:magneticWell}
\end{figure}

Neutrons with their magnetic moment oriented antiparallel to a magnetic field are decelerated by an increasing magnetic field amplitude, and are called low field seekers (LFS). Neutrons having a magnetic moment parallel to the magnetic field are accelerated and are called high field seekers (HFS). The storage volume of \tspect is a three-dimensional magnetic potential well for LFS, depicted in Fig.~\ref{fig:magneticWell}. All magnetic field sources are simulated externally using the open source python package \magpylib~\cite{ortnerMagpylibFreePython2020}. For each source, a mesh is computed containing the field components and derivatives relevant for the tricubic interpolation integrated in \pentrack~\cite{lekienTricubicInterpolationThree2005}. For that purpose, we extended \pentrack for accepting external tricubic components as parameters.

\subsubsection{Superconducting coils}

The longitudinal field is produced by a series of superconducting coils in a cryostat. It features two characteristic peaks ($x\simeq 0\,\textrm{mm}$ and $x\simeq 2200\,\textrm{mm}$) which confine UCNs longitudinally. The global amplitude of the field can be shifted by varying the electric current in the coils (see Appendix~\ref{sec:coils}). In addition, two independent \emph{gradient} coils are used to change the slope of the magnetic field around $x\simeq 0\,\textrm{mm}$ (see Fig.~\ref{fig:tspectExperiment}), which is relevant for the SF of UCNs discussed in Sec.~\ref{sec:sf}. The simulation framework incorporates those various coil settings. The iron yoke surrounding the cryostat reduces the fringe field outside. It is not included in the simulation framework, but it has been shown that the shape and the high homogeneity of the magnetic fields at the center of the cryostat is unaffected by the magnetic flux return yoke~\cite{konradMagneticShieldingNeutron2014}.

\subsubsection{Halbach octupole}
\label{sec:Halbachoctupole}

The transversal field is produced by a Halbach array made from permanent magnets in an octupole configuration installed inside the cryostat~\cite{aulerSPECTSpinflipLoaded2024}. The octupole is $1.38 \,\textrm{m}$ long, segmented into 24 rings with an inner radius of $R_{i}=54\, \textrm{mm}$ and outer radius of $R_{o} = 84.1 \,\textrm{mm}$. Each ring is formed of 32 $\textrm{Sm}_{2} \textrm{Co}_{17}$ segments, with a nominal remanence of $1.1\,\textrm{T}$ and a relative permeability of $\mu_{r} = 1.1$.

\begin{figure}[th]
  \centering
  \includegraphics[width=0.49\textwidth, trim={0 0.2cm 0 0cm},clip]{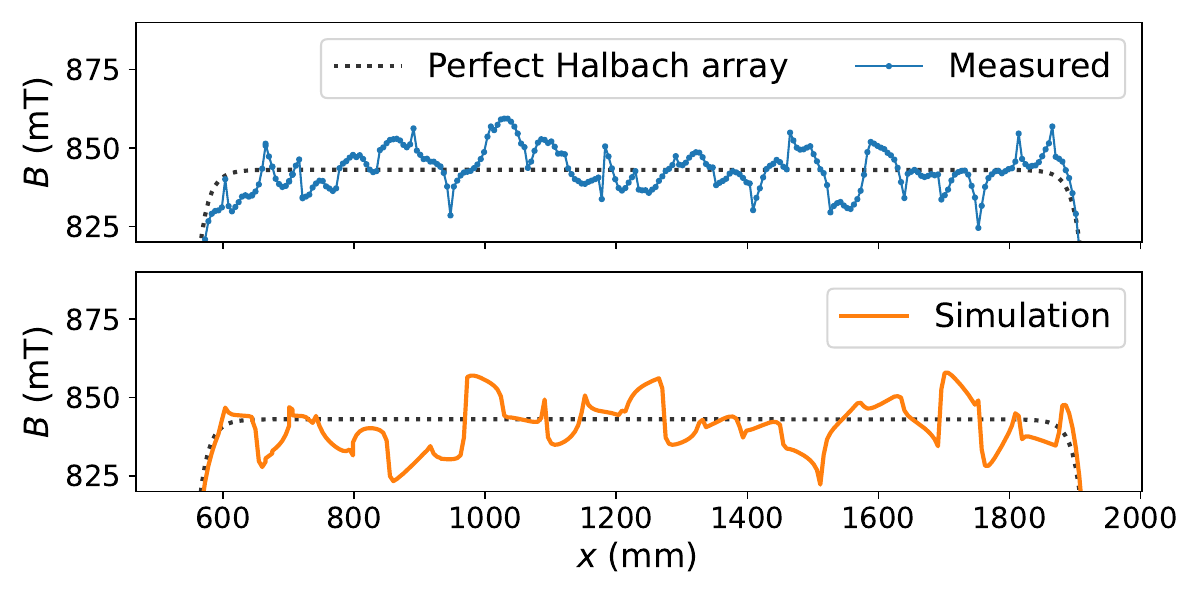}
  \captionsetup{width=.9\linewidth}
  \caption{Magnetic field amplitude $|\vec B|$ of the octupole along the longitudinal direction $x$. The scan is done at a fixed angle, and a radius of $52.5 \, \textrm{mm}$. For clarity, uncertainty ranges are not displayed.}
  \label{fig:octupole_dataVSsimu}
\end{figure}

In order to mimic small misalignments inherent to the Halbach array construction, each simulated segment undergoes a small random Gaussian perturbation of its magnetization orientation and amplitude, its position, and its dimensions. Those perturbations have been tuned in the simulation code to match the patterns observed during measurements of the octupole magnetic field\footnote{Reanalysis of \cite{haackVollmagnetischeSpeicherungUltrakalten2016}}, as shown in Fig.~\ref{fig:octupole_dataVSsimu}. The standard deviation of the perturbations were chosen as $11 \,\textrm{mT}$ and $2 \,\textrm{°}$ for the magnetization amplitude and orientation, $0.03 \,\textrm{mm}$ for the position, $0.2 \,\textrm{°}$ for the orientation, and $0.2 \,\%$ for the dimension of each segment.

\subsubsection{Trap energy potential}

The total energy $E$ of the UCNs in the trapping volume must be below the potential depth $V_{\textrm{trap}}$ so that they are confined fully magnetically. $V_{\textrm{trap}}$ is limited by the octupole field, as its amplitude near its surface is lower than the two longitudinal field peaks formed by the superconducting coils. Therefore, the trapping potential corresponds to the minimum value of the sum of the neutron magnetic and gravitational potential energies computed at the octupole surface. From the simulated octupole, we find
\begin{align}
  V_{\textrm{trap}} = \left.\min(V_{\textrm{magnetic}} + V_{\textrm{gravity}}) \right|_{r=R_{i}} \simeq 48\,\textrm{neV},
\end{align}
where
\begin{align}
  &V_{\textrm{gravity}}(x,y,z) \simeq z\cdot102.4 \,\textrm{neV/m},\nonumber\\
  &V_{\textrm{magnetic}}(x,y,z) \simeq |B(x,y,z)| \cdot 60.3 \, \textrm{neV/T},\nonumber
\end{align}
and $B$ is the combined magnetic field of the superconducting coils and the octupole. The exact value of $V_{\textrm{trap}}$ in \tspect is unknown. The minimum magnetic field amplitude of the trap is at its center $(x,y,z) = (1.26\,\textrm{m}, 0, 0)$, with a value of about \SI{0.21}{\tesla}, corresponding to a minimum magnetic potential for LFS of $V_{\textrm{min}} = \min(V_{\textrm{magnetic}}) \simeq 12.5 \, \textrm{neV}$.

In practice, some orbits allow UCNs with higher energy than $V_{\textrm{trap}}$ to remain in the trapping volume for durations of the same order as the free neutron lifetime $\tau_{n}$, before leaking out. Those so-called marginally trapped neutrons are one of the main sources of systematic uncertainties for UCN lifetime bottle experiments, as the loss rate related to their escape of the trap adds up to the decay rate of the neutron.

In order to remove this source of bias, \tspect currently relies on a \emph{cleaning} method for removing marginally trapped UCNs. After neutrons are filled into the trap, and before the storage period begins, the in-situ neutron detector is partially inserted in the trapping volume, but remains in the slope of the second magnetic peak (see Fig.~\ref{fig:magneticWell}). Ideally, mostly marginally trapped neutrons, thus having an energy higher than fully storable UCNs, can reach it and be absorbed on its surface.

\begin{figure}[ht]
  \centering
  \includegraphics[width=0.39\textwidth, trim={0 0cm 0 0cm},clip]{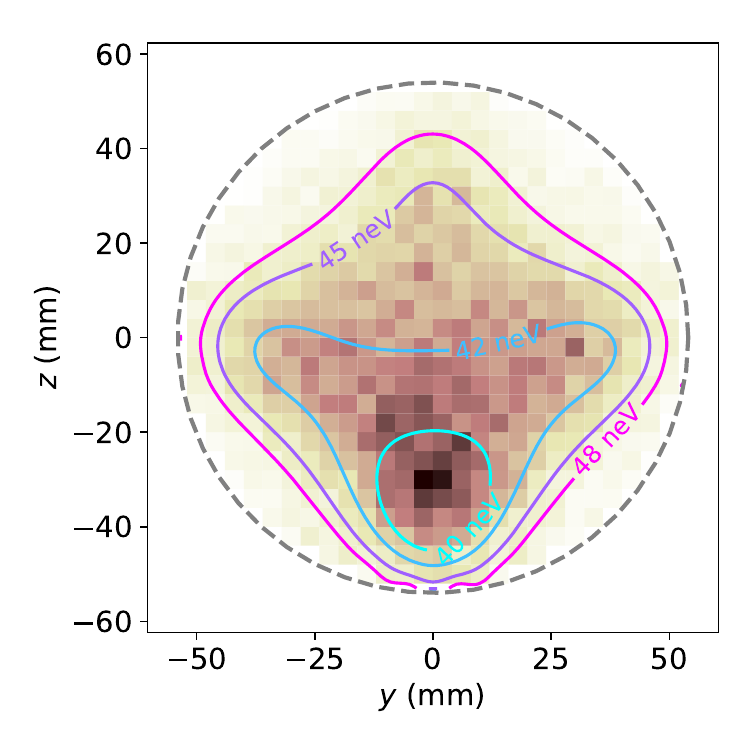}
  \captionsetup{width=.9\linewidth}
  \caption{Spatial distribution of the UCN detection on the \tspect detector during a cleaning phase at $x_{\rm clean} = 1850 \, \textrm{mm}$, derived from simulations. The equipotential lines are computed from the combination of magnetic and gravitational potential energy.}
  \label{fig:cutplanex1.8m}
\end{figure}

As shown in Fig.~\ref{fig:cutplanex1.8m}, the spatial distribution of the UCN detection derived from simulations is in agreement with the potential energy computed on the surface of the detector plate, for a simulated cleaning position of $x_{\textrm{clean}}=1850\,\textrm{mm}$.

\begin{figure}[ht]
  \centering
  \includegraphics[width=0.49\textwidth, trim={0 0.5cm 0 0cm},clip]{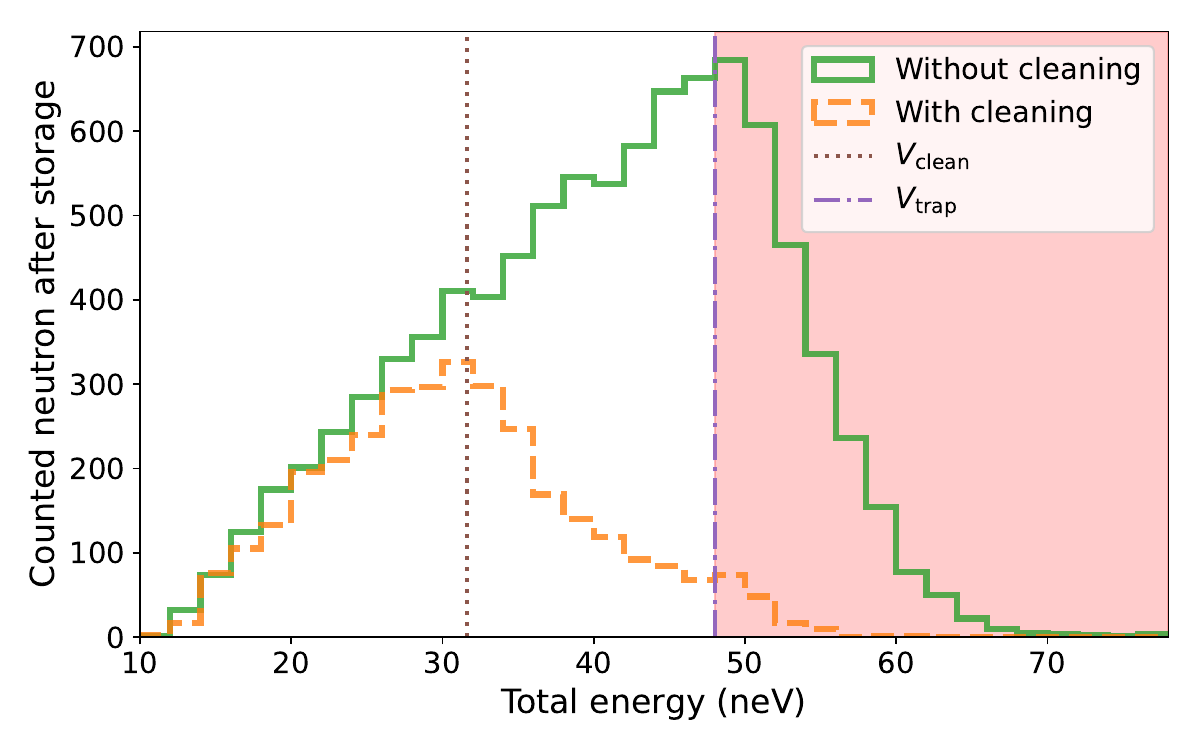}
  \captionsetup{width=.9\linewidth}
  \caption{Total energy spectrum of detected neutrons. \tspect is at a height of $59 \,\textrm{cm}$, UCNs are filled into the trap until $t=25 \,\textrm{s}$, and stored for $50 \,\textrm{s}$ before being counted by the detector. For the cleaning simulation, the detector is positioned at $x=1820 \,\textrm{mm}$ until $t=100 \,\textrm{s}$ before being retracted for the storing phase. Marginally trapped neutrons have energies above $V_{{\textrm{trap}}}=\SI{48}{\nano\electronvolt}$ (filled region).}
  \label{fig:countedHdistribution}
\end{figure}

Simulations show that a fraction of the marginally trapped UCN population keeps a predominantly transverse momentum, and therefore lack longitudinal momentum to reach the detector. To remove this marginally trapped population, the absorbing surface of the detector can be moved deeper into the trap, at the cost of removing a significant fraction of the storable population as well, as shown in Fig.~\ref{fig:countedHdistribution}. Based on simulations, improvements of the cleaning technique for sharpening the cut in the energy spectrum are still being investigated.

\subsection{UCN filling and spin-flipping units}
\label{sec:sf}

\begin{figure}[th]
  \centering
  \includegraphics[width=0.49\textwidth, trim={0cm 0cm 0cm 0cm},clip]{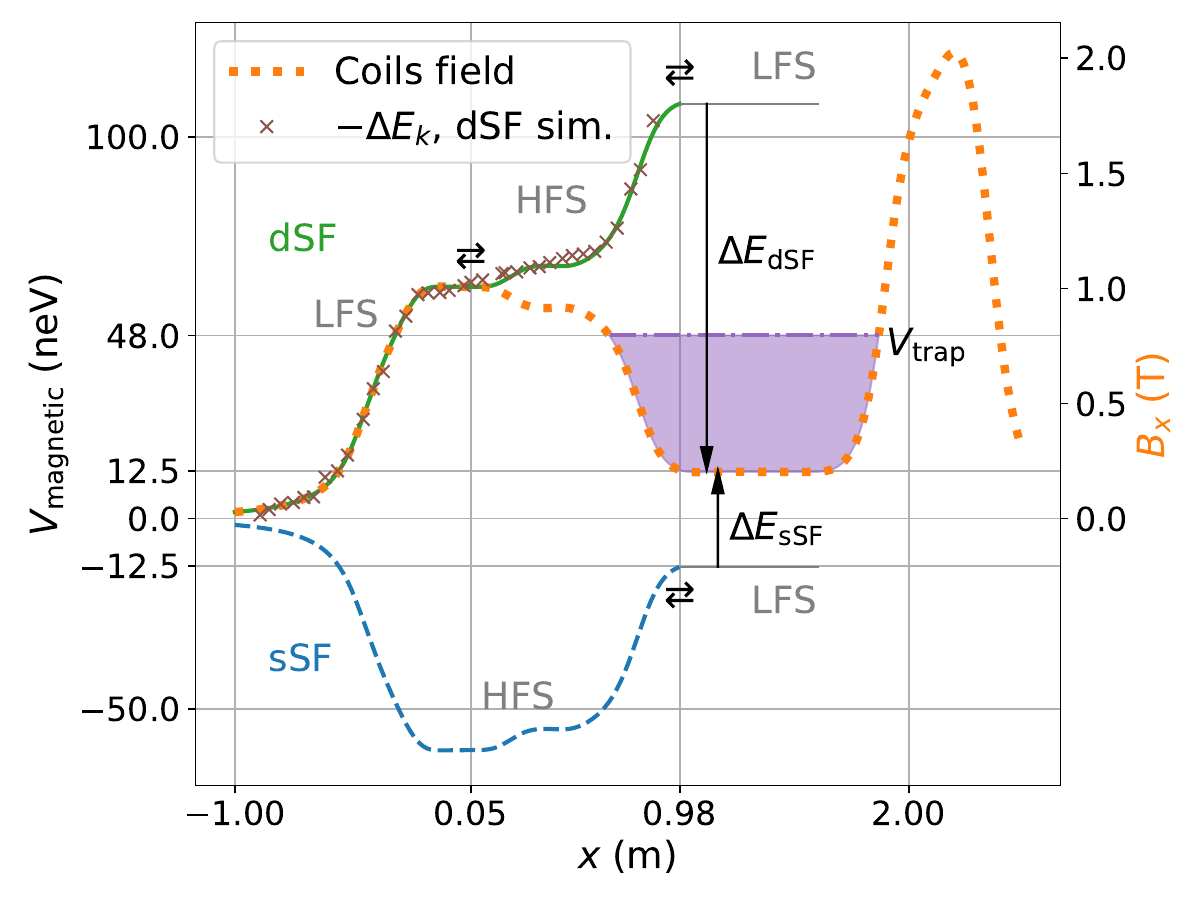}
  \captionsetup{width=.9\linewidth}
  \caption{Magnetic potential curves (solid green for dSF, and dashed blue for sSF) seen by UCNs as they enter the \tspect trap (filled purple) from the left, and get spin-flipped ($\rightleftarrows$). Dotted orange: longitudinal coil magnetic field amplitude. sSF UCNs, following the HFS$\to$LFS bottom path, undergo an increase of total energy of $\Delta E_{\textrm{sSF}}\simeq+25\,\textrm{neV}$. dSF UCNs, following the LFS$\to$HFS$\to$LFS top path, undergo a reduction of total energy of $\Delta E_{\textrm{dSF}}\simeq -96 \,\textrm{neV}$. The brown crosses correspond to the opposite of the kinetic energy shift, $-\Delta E_k$, sampled along the UCN trajectories for a double spin-flip simulation (see discussion in the text).}
  \label{fig:SFpotentialsCurves}
\end{figure}

On the path to the storage volume, neutrons undergo a kinetic energy shift due to the magnetic amplitude variation in the longitudinal direction, whose sign depends on their polarization state (HFS or LFS). In order for UCNs to be storable, they must be in the LFS state in the trapping region, and have a total energy below $V_{\textrm{ trap}}$. This is achieved by using either one or both adiabatic spin flipping units (SFUs)~\cite{abragamPrinciplesNuclearMagnetism1961, luschikovCalculationNeutronAdiabatic1984} installed along the neutron path (see Fig.~\ref{fig:tspectExperiment}). Their purpose is to reverse the neutron polarization at two defined longitudinal positions, $x_{\textrm{SF1}}=50\,\textrm{mm}$ and $x_{\textrm{SF2}}=980\,\textrm{mm}$. When passing a SFU, a LFS is transformed into HFS, and vice versa. Fig.~\ref{fig:SFpotentialsCurves} summarizes the approximate potential energy shift experienced by neutrons for two SF setups:
\paragraph{Single spin-flip}

After passing the first magnetic peak, HFS are spin-flipped around $x_{\textrm{SF2}}$ into LFS, making them storable. This technique is referred to as the single spin-flip (sSF) technique. Because the magnetic potential energy of those HFS, $\left. V_{\textrm{magnetic}}\right|_{x_{\textrm{SF2}}}\simeq -12.5\,\textrm{neV}$, is reversed once spin-flipped into LFS, this method increases those UCNs total energy by
\begin{align}
  \Delta E_{\textrm{sSF}} = 2\left| V_{\textrm{magnetic}}\right|_{x_{\textrm{SF2}}} \simeq +25\,\textrm{neV}.
\end{align}
    
For these neutrons to be fully storable, their total energy range allowed before being spin-flipped must therefore be
\begin{align}
  E \in [0,  V_{\textrm{trap}} - \Delta E_{\textrm{sSF}}] \simeq [0, 23]\,\textrm{neV}.
\end{align}

\paragraph{Double spin-flip}

In addition, another SFU can be activated, located at $x_{\textrm{SF1}}$, in the middle of the first magnetic field peak. In that case, initially LFS UCNs are slowed down before reaching SF1, get spin-flipped into HFS, and continue to be slowed down as they reach SF2. This technique is referred to as the double spin-flip (dSF), and reduces those UCNs total energy by
\begin{align}
  \Delta E_{\textrm{dSF}} = -2\left| V_{\textrm{magn}}\right|_{x_{\textrm{SF1}}} + 2\left| V_{\textrm{magn}}\right|_{x_{\textrm{SF2}}}  \simeq -96 \,\textrm{neV}.
\end{align}
The total energy range (before SF1) allowed for being storable is here
\begin{align}
  E \in [V_{\textrm{min}}-\Delta E_{\textrm{dSF}}, V_{\textrm{trap}} - \Delta E_{\textrm{dSF}}] \simeq [108, 144]\,\textrm{ner  V}.
\end{align}

Each SF scheme allows for storing UCNs extracted from a different part of the energy spectrum. Combined with \tspect height variations, the neutron loading can be optimized to maximize the number of confined UCNs.

In addition to the theoretical calculations based on the magnetic field potentials shown in Fig.~\ref{fig:SFpotentialsCurves}, we also display the UCN kinetic energy shift sampled from a dSF simulation. It is computed as the difference between the neutron kinetic energy at the longitudinal position $x$ and the position $x=\SI{-2}{\meter}$ (i.e., before entering \tspect), $\Delta E_{k}(x) = E_k(x) - E_k(x=\SI{-2}{\meter})$. The simulations confirm the theoretical predictions $\Delta V_{\textrm{magnetic}} \simeq -\Delta E_{k}$.


\subsubsection{Spin-flip simulations}

\begin{figure}[ht]
  \centering
\includegraphics[width=0.3\textwidth, trim={5.1cm 10.4cm 2cm 9cm},clip]{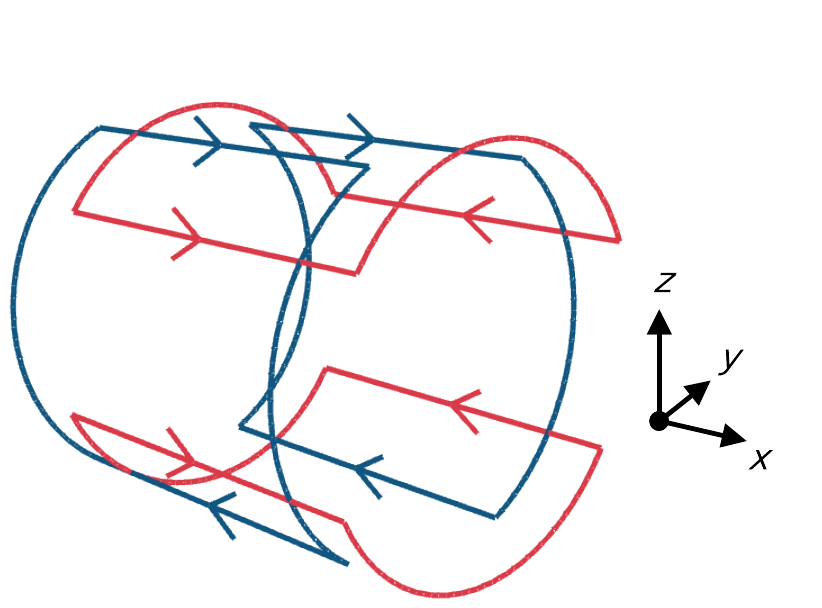}
  \includegraphics[width=0.49\textwidth, trim={0 -1cm 0 0cm},clip]{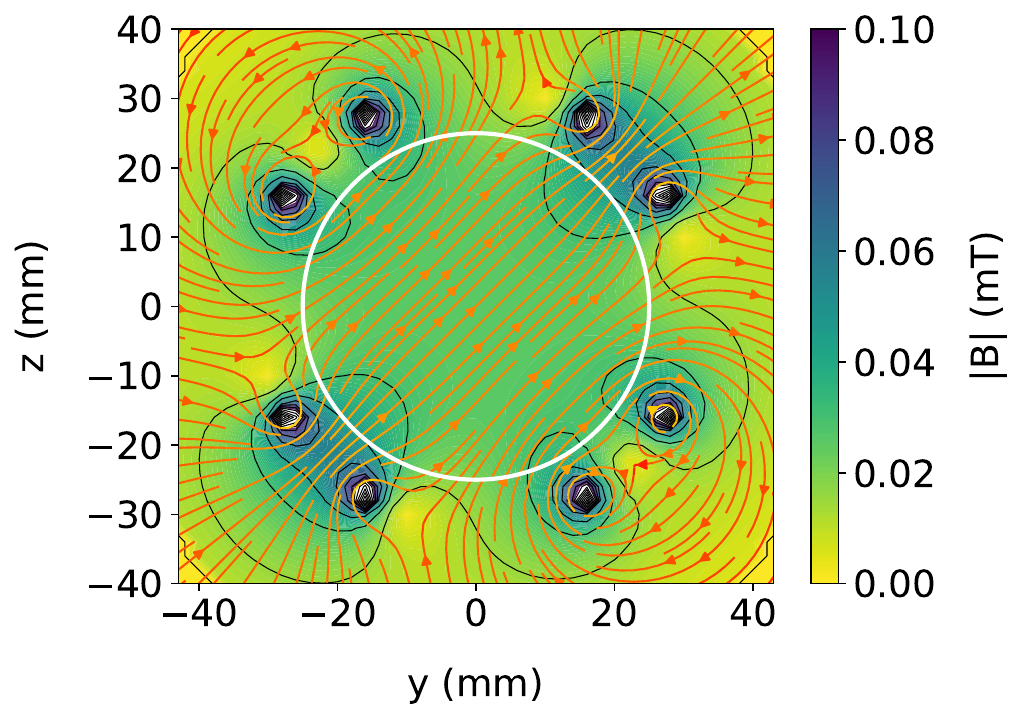}
  \captionsetup{width=.9\linewidth}
  \caption{Top: 3D view of the two saddle coils for the second spin-flipper unit. The arrows indicate the RF electrical current circulation at a given time. Bottom: simulated magnetic field amplitude $|\vec B|$ and field lines the $y$-$z$ plane at the center of the SFU. As the current oscillates at the RF, the orientation of the field rotates about the central $x$ axis. The UCN guide tube boundary is indicated by the white circle. The 8 points on the plane trough which the coil wires pass are visible in black.}
  \label{fig:SFmagneticField}
\end{figure}

Both SFUs currently consist of two orthogonal saddle coils, fed by RF current with a $\pi/2$ phase offset. This produces a homogeneous magnetic field in the $y$-$z$ plane of the UCN guides, that rotates around the longitudinal $x$ axis, as shown in Fig.~\ref{fig:SFmagneticField}. By matching the RF with the neutron spin precession frequency, their spin can be manipulated and ultimately adiabatically reversed as they exit the SFUs~\cite{abragamPrinciplesNuclearMagnetism1961, luschikovCalculationNeutronAdiabatic1984}. After a defined filling time, the SFU and guide tube assembly is retracted from the storing region, leaving UCNs in a material free environment.

Because the second SFU position is in the octupole field region during the filling process, it is integrated into a compensation Halbach octupole which counters the storage octupole, ensuring that the magnetic field gradient is dominant in the longitudinal direction. The compensation octupole is implemented in the simulation framework similarly as for the storage octupole field, discussed in Sec.~\ref{sec:Halbachoctupole}. As it is attached to the SFU translation assembly, and because \pentrack does not yet implement moving magnetic fields, the amplitude of the simulated compensation octupole field is simply scaled down to zero once the filling phase ends.

\begin{figure}[ht]
  \centering
  \includegraphics[width=0.49\textwidth, trim={0cm 0cm 0cm 0cm},clip]{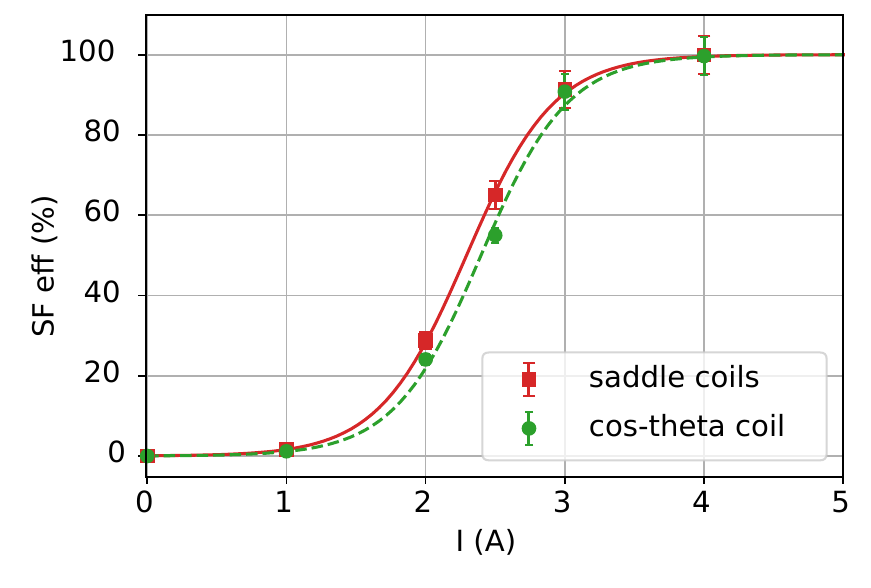}
  \captionsetup{width=.9\linewidth}
  \caption{Spin-flip efficiency from simulations of the SFU located at $x_{\textrm{SF2}}$ depending on the coil input current $I$. Here, two coil designs are compared, a pair of saddle coils, as pictured in Fig.~\ref{fig:SFmagneticField}, and a single cos-theta coil with four loops.}
  \label{fig:SFefficiency}
\end{figure}

Because spin-tracking in \pentrack is computationally expensive, dedicated simulations have been developed for characterizing the SFU efficiency only. A series of longitudinal current segments, approximating the SFU coil paths, are being fed by RF current. The initial neutron parameters like velocity and position before entering the SFU region are first sampled. Multiple SFU coil designs can be tested and optimized, by counting the number of neutron successfully spin-flipped, the SF efficiency can be deduced, as shown in Fig.~\ref{fig:SFefficiency}. By varying the amplitude of the current, various magnetic field configurations from the coils of the cryostat (see Appendix~\ref{sec:coils}) can also be tested.

Future studies employing spin-flip (SF) simulations will assess SF efficiency as a function of UCN incoming position, velocity, direction, magnetic field gradient, and SF coil current. These calculated SF probabilities can then be incorporated directly into our full measurement cycle simulations, eliminating the need for explicit spin tracking during the spin-flip process. The results of these full simulations can then be compared with measurement data from the \tspect detector.



\section{End-to-end simulation}
\label{sec:endtoendsimulation}

\subsection{Configurations}

All subsystems introduced in previous sections have been assembled into a full simulation setup. The framework incorporates the most relevant degrees of freedom for setting the experiment (height relative to the beamport, SF technique, SF and detector assembly positions and speed, timing, field amplitude, beamline geometry, ...). For this purpose, a dedicated companion software, \penconf, has been developed. Written in Python, it offers an additional configuration level to \pentrack. Custom experimental settings can be assigned to variables, for example, the UCN filling and storage times, cleaning duration and position, etc. For each new configuration, \penconf can then automatically generate a \pentrack configuration file, eliminating the need to manually write it. With \penconf, a pre-configuration must first be constructed, where all the possible geometries, fields, and UCN sources are defined. CAD volumes can be grouped together as being part of a single geometrical entity which have multiple states (for example, a shutter, that can be closed or open). At the time of developing the framework, moving geometries were not yet implemented in \pentrack. However, volumes can be (de)activated for periods of time, and we therefore currently rely on that feature to mimic their movement and change of state. From a CAD software, we first export each component into several predefined positions, which are then loaded into \pentrack as separate volumes, which are be alternatively (de)activated during the simulation. The automation of this complex task is also carried out by \penconf. The impact of this moving geometries approximation will be studied in the future, once Dopper-shift will be implemented in the simulation engine. We expect minor effects from the moving spin-flipper assembly ($24 \textrm{mm/s}$), detector surface ($10 \textrm{mm/s}$), or UCN shutter.

The framework can thus handle end-to-end simulations of UCNs produced by the PSI source, guided in the beamline, up to the filling, storage and ultimately detection inside the trapping region. For those full simulations, the process of neutron spin-flip is carried out by surfaces in the $y$-$z$ plane, positioned at both SFU positions, $x_{\textrm{SF1}}$ and $x_{\textrm{SF2}}$, that instantaneously reverse the polarization of UCNs going through them. Various spin-flip probabilities can be attributed to those surfaces, which thus reproduces the non 100\% efficiency of the actual SFUs.

\subsection{Measurement run}

\begin{figure}[ht]
  \centering
  \includegraphics[width=0.49\textwidth,  trim={0cm 0cm 0cm 0cm},clip]{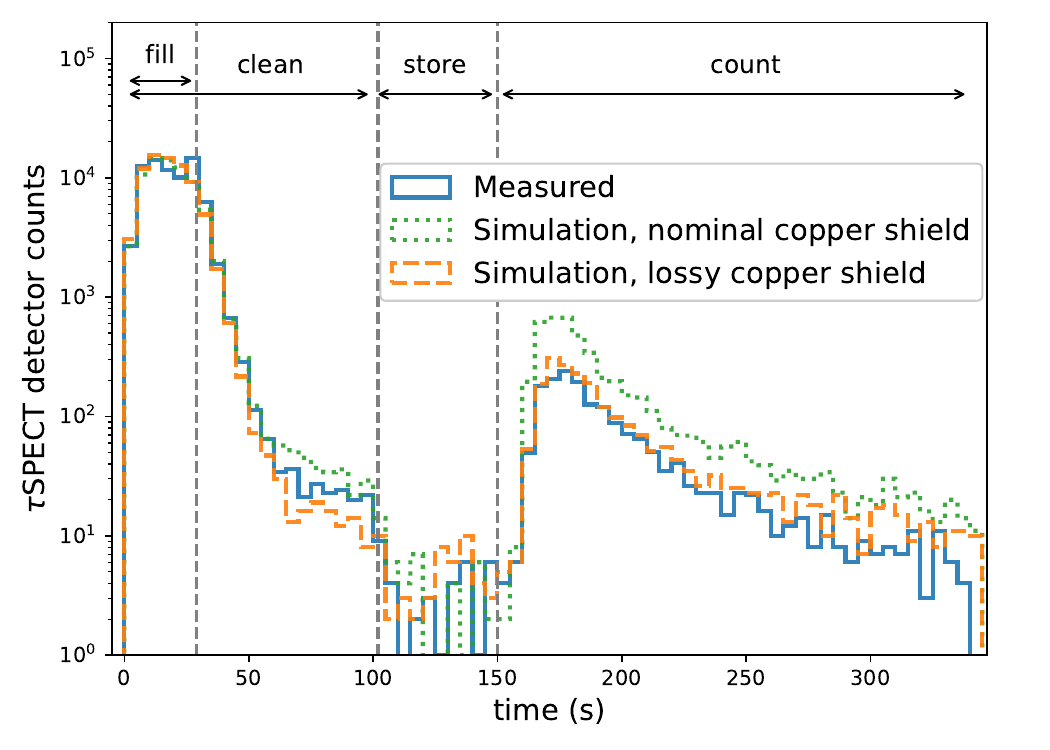}
  \captionsetup{width=.9\linewidth}
  \caption{Comparison between measurement and simulations of \tspect main detector counts during a single measurement cycle consisting of $27\,\textrm s$ of filling, a cleaning from $0$ to $100\,\textrm s$, followed by $50\,\textrm s$ of storage, and finally $200\,\textrm s$ of counting. See Appendix~\ref{sec:materials} for the nominal and lossy copper shield material properties.}
  \label{fig:fullSim}
\end{figure}

A characteristic counting spectrum of the main detector of \tspect for a single storage measurement cycle is shown in Fig.~\ref{fig:fullSim}. It consists of mainly four phases.

\begin{itemize}
    
\item {Filling: } The UCN shutter is open, allowing UCNs to enter \tspect. The SFUs assembly is already inserted in the trap and active. During that period, the detector is at a cleaning position, and counting UCNs. Many neutrons are detected, mostly high energetic ones, flying through the trap.
\item {Cleaning: } The UCN shutter is closed, the SFUs are deactivated, and the SFU assembly is removed from the trap. The detector keeps its cleaning position for removing marginally trapped neutrons. A few remaining high energetic UCNs are still pouring from the guides into the trap, and get detected as well.
\item {Storing: } The detector is retracted from its cleaning position, leaving UCNs in a material free environment and confined solely by the magnetic field. At that stage, mostly background signals are detected.
\item {Counting: } The detector is pushed deep inside the trap, removing and counting the remaining UCNs.

\end{itemize}

The shape of the counting spectrum of the \tspect main detector depends on the whole simulations settings: the initial conditions of the UCNs, the material properties of all the geometries, and the magnetic field configuration. One can appreciate the accuracy of the simulation framework in reproducing the counting spectrum. A Poisson background of $0.9$ count per second is added to the simulation spectrum, close to what is measured.

With nominal material properties of \tspect geometries, the simulations predict almost three times more counted (and stored) UCNs than what is measured. This is visible as a higher counting peak amplitude (green dotted line in Fig.~\ref{fig:fullSim}) compared to the measured spectrum (solid blue). In order for the simulation to match the observed amplitude (dashed orange), the loss per bounce of the copper shield (Fig.~\ref{fig:tspectExperiment}) had to be drastically increased compared to the expected nominal UCN property of copper (see Appendix~\ref{sec:materials}). Indeed, since the copper shield is the main geometry in contact with trapped UCNs during the filling phase, its material properties can influence the total number of stored and thus counted UCNs after the storage phase. The UCNs detected during the filling phase are high energetic ones that fly straight through the trap, and are not in contact with the copper shield. Hence, the filling peak is not impacted by its material property. Although it is not yet clear whether the copper shield is to blame, this calls for a future inspection of any loss mechanisms to explain this discrepancy between measurements and simulations. In 2025, after re-polishing the copper shield, new measurements will be compared again with simulations to clarify this point.

\begin{figure}[ht]
  \centering
    \includegraphics[width=0.49\textwidth,  trim={0cm 0cm 0cm 0cm},clip]{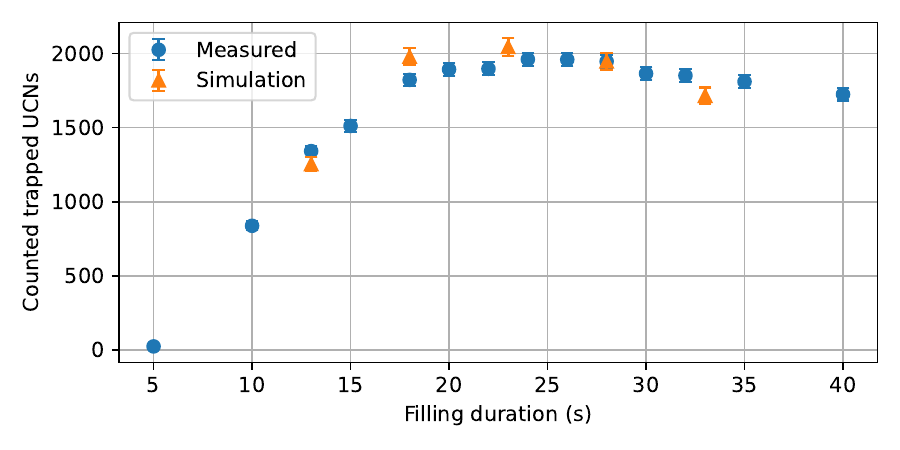}
  \captionsetup{width=.9\linewidth}
  \caption{Comparison between simulation and measurement at PSI of a filling time parameter scan using the dSF technique at a height of $59\,\textrm{cm}$. The cleaning phase is applied from $t=0\,\textrm{s}$ to $t=100\,\textrm{s}$ at a position $x_{\textrm{clean}}=1800\,\textrm{mm}$, followed by a $50\,\textrm{s}$ storage.}
  \label{fig:fillingTimeScan}
\end{figure}

As a more advanced example, Fig.~\ref{fig:fillingTimeScan} shows a filling time scan performed at PSI in 2024, and a comparison with simulations. Both simulation and measurement agree on an optimal filling time of about $25\,\textrm{s}$.


\newpage
\section{Conclusion}
\label{sec:conclusion}

In this paper, we have introduced a comprehensive UCN simulation framework tailored for the neutron lifetime experiment \tspect. This framework has been crucial in characterizing the experiment and has yielded results consistent with preliminary analyses from measurements conducted by \tspect at PSI in 2024. We demonstrated that the framework effectively handles complex features, such as realistic magnetic fields and moving geometries, within a unified structure built on the Monte Carlo simulation software \pentrack. Additionally, we developed two companion tools, \penconf and \penplot, designed to streamline the configuration and analysis of \pentrack simulations. These tools are anticipated to significantly aid the development of similar frameworks for other UCN experiments. A clean version of the simulation framework itself will be made public on GitLab in the near future.

The simulations have not only validated the UCN storage design of \tspect but have also provided valuable insights for guiding future improvements and identifying systematic uncertainties associated with the free neutron lifetime measurement of the current setup. For example, techniques to enhance neutron loading efficiency or to mitigate effects of marginally trapped UCNs can be evaluated through simulations prior to hardware implementation.

During the second half of 2024 and 2025, \tspect underwent an extensive commissioning phase, during which various parameter scans were performed. Based on those measurements, we plan to fine-tune the simulation framework using large-scale computer grid to increase statistics.

During the second half of 2024 and throughout 2025, the \tspect experiment underwent an extensive commissioning phases involving detailed parameter scans. Leveraging these measurements, we plan to fine-tune the simulation framework using large-scale computing grids to significantly enhance statistics.  Systematic studies will focus on several non-exhaustive aspects, beginning with poorly constrained UCN surface loss parameters, and extending the characterization of the  marginally trapped UCN population, including their cleaning efficiency, and phase-space evolution and randomization. We will further investigate the detector surface properties such as its Fermi potential and roughness, alongside counting window effects like incomplete detection during finite windows or background accumulation during extended counting periods. Source-related effect, particularly fluctuations in initial UCN numbers and energy spectrum drift, will be studied as well. Magnetic trap performance will be inspected through depolarization rates, spin-flip losses, field inhomogeneities, and adiabaticity violations.

\newpage
\subsection*{Acknowledgements}

This work has been supported by the Cluster of Excellence Precision Physics, Fundamental Interactions, and Structure of Matter (PRISMA+ EXC 2118/1) funded by the German Research Foundation (DFG) within the German Excellence Strategy (Project ID 39083149).

We thank Dr. Georg Bison for providing the CAD model of the PSI UCN source, and Dr. Geza Zsigmond for the useful discussions and for providing \mcucn comparison data.


\subsection*{Author contributions}

\textbf{S.V.} built the simulation framework, wrote the software packages, performed the main analysis, and wrote this document;
\textbf{N.P.} performed the spin-flipping simulations and analysis;
\textbf{U.B.} tested the simulation and analysis software;
\textbf{J.A., M.E., V.E., M.F., K.F., S.K., B.L., N.P., D.R., S.V., and N.Y.} contributed to the reassembly and data taking of \tspect at PSI;
All authors discussed the results and commented on the
manuscript.

\newpage
\appendix
\section{Appendix}
\subsection{Simulation material properties}

\label{sec:materials}

\begin{table}[!ht]
\begin{center}
  \resizebox{0.49\textwidth}{!}{\begin{tabular}{ |l|l|l|l|l| } 
    \hline
    Material & $V$ (neV) & $W$(neV)& $P_{\textrm{Lamb.}}$ & $P_{\textrm{SF}}$ \\ \hline 
    $\textrm{AlMg}_{3}$ & 54 & 0.0054 & 0.1 & $10^{-5}$\\
    NiMo & 220 & 0.0660 & 0.05 & $2\cdot 10^{-6}$ \\ 
    DLC & 230 & 0.253 & 0.5 & $2\cdot 10^{-6}$\\ 
    Polished steel & 174 & 0.0592 & 0.4 & $10^{-5}$ \\
    Steel loss & 174 & 1 & 0.4  & $10^{-5}$ \\
    Steel loss 100 & 174 & 100 & 0.4  & $10^{-5}$ \\
    $^{58}\textrm{NiMo (85:15)}$   & 308 & 0.02755 & 0.05 & $2\cdot 10^{-6}$  \\
    $\textrm{Sm}_{2}\textrm{Co}_{17}$ & 44 & 0.01 & 0.1 & $10^{-5}$ \\
    Cu (\textbf{nominal}) & 171 & 0.0726 & 0.2 & $10^{-5}$ \\
    Cu W70 (\textbf{lossy}) & 171 & 70 & 0.2 & $10^{-5}$ \\
    Quartz & 103 & 0.038 & 0.2 & $2\cdot 10^{-6}$ \\
    Spin flipper & 0 & 0 & 0 & 1 \\
    UCN det V0 & $10^{-5}$ & 1 & 0 & 0 \\ 
    UCN det V10 & $10$ & 1 & 0 & 0 \\
    \hline
  \end{tabular}
  }
\end{center}
\captionsetup{width=0.9\linewidth}
\caption{Material properties. $V$ and $W$ refer to the real (Fermi) and imaginary part of the potential. $P_{\textrm{Lamb.}}$ and $P_{\textrm{SF}}$ respectively refer to the probability of non-specular (Lambertian) reflection, and of spin-flipping}
  \label{tab:material}
\end{table}

\begin{table}[!ht]
\begin{center}
  \begin{tabular}{ |l|l| } 
    \hline
    Geometry & Material\\ \hline 
    Aluminum lid & $\textrm{AlMg}_{3}$\\ 
    Vertical guide & NiMo \\ 
    Storage vessel & DLC \\ 
    Central tube & DLC \\ 
    Cryopump & UCN det \\ 
    Storage flaps & DLC \\ 
    NLK flap & DLC \\ 
    Source guides & NiMo \\ 
    AlMg3 foil grid & Polished steel \\ 
    AlMg3 foil & $\textrm{AlMg}_{3}$ \\
    \hline
  \end{tabular}
\end{center}
\captionsetup{width=.9\linewidth}
\caption{Materials used for PSI UCN source geometries.}
  \label{tab:PSImaterial}
\end{table}

\begin{table}[!ht]
\begin{center}
  \begin{tabular}{ |l|l| } 
    \hline
    Geometry & Material\\ \hline 
    Reduction flange & Polished steel\\ 
    Beamline guides & Polished steel\\ 
    Beamline UCN mirror & $^{58}\textrm{NiMo (85:15)}$   \\ 
    Monitoring detector flange & Polished steel \\ 
    Monitoring detector surface & UCN det V0\\
    Quartz guide 1st part & Cu \\
    Quartz guide 2nd part & Quartz \\
    SF surfaces & Spin flipper \\
    Halbach segments & $\textrm{Sm}_{2}\textrm{Co}_{17}$ \\
    UCN shutter, source side & Steel loss \\
    UCN shutter, \tspect side & Steel loss 100 \\
    Copper junction & Cu \\
    Copper shield & Cu W70 \\
    \tspect detector surface & UCN det V10 \\
    \hline
  \end{tabular}
\end{center}
\captionsetup{width=.9\linewidth}
\caption{Materials used for \tspect geometries.}
  \label{tab:tspectmaterial}
\end{table}

\newpage
\subsection{Coils}
\label{sec:coils}

\begin{table}[!ht]
  \begin{center}
    
    \resizebox{0.49\textwidth}{!}{
      \begin{tabular}{|l|l|l|l|l|l|}
      \hline
      Coil &  $r_{o} - r_{i}$ (m)& $x_f - x_{i}$ (m) & $r_{i}$  & $x_i$ (m) & Current (A) \\
      \hline
      1 & 0.0048 & 0.2866 & 0.129 & -0.4294 & $256.53 \cdot I_{\textrm{main}}$ \\
      \hline
      2a & 0.016 & 0.0715 & 0.2548 & -0.137 & $146.85 \cdot I_{\textrm{main}}$ \\
      \hline
      2b & 0.007 & 0.0715 & 0.2708 & -0.137 & $189.81 \cdot I_{\textrm{main}}$ \\
      \hline
      2c & 0.02622 & 0.0715 & 0.2778 & -0.137 & $256.46 \cdot I_{\textrm{main}}$ \\
      \hline
      4a & 0.0176 & 0.0468 & 0.2428 & 0.0673 & $146.90 \cdot I_{\textrm{main}}$ \\
      \hline
      4b & 0.0126 & 0.0468 & 0.2604 & 0.0673 & $189.26 \cdot I_{\textrm{main}}$ \\
      \hline
      4c & 0.0348 & 0.0468 & 0.273 & 0.0673 & $256.41 \cdot I_{\textrm{main}}$ \\
      \hline
      6 & 0.0048 & 0.1397 & 0.1408 & 0.1396 & $256.41 \cdot I_{\textrm{main}}$ \\
      \hline
      7 & 0.0078 & 0.1781 & 0.15 & 0.2954 & $256.50 \cdot I_{\textrm{main}}$ \\
      \hline
      8 & 0.009 & 0.3159 & 0.15 & 0.4741 & $256.41 \cdot I_{\textrm{main}}$ \\
      \hline
      9 & 0.0018 & 0.6689 & 0.1928 & 0.9906 & $256.41 \cdot I_{\textrm{main}}$ \\
      \hline
      10 & 0.0198 & 0.3029 & 0.12 & 1.9 & $256.39 \cdot I_{\textrm{main}}$ \\
      \hline
      11 & 0.027 & 0.1502 & 0.12 & 2.2029 & $256.32 \cdot I_{\textrm{main}}$ \\
      \hline
      3 & 0.0042 & 0.00455 & 0.1312 & -0.1178 & $256.41 \cdot I_{\textrm{c3}}$ \\
      \hline
      5 & 0.0042 & 0.00455 & 0.1312 & 0.1132 & $256.41 \cdot I_{\textrm{c5}}$ \\
      \hline
      \end{tabular}
      }
  \end{center}
  \captionsetup{width=.9\linewidth}
  \caption{Coil properties used to generate longitudinal field in \tspect~\cite{konradMagneticShieldingNeutron2014}. $r_i$ and $r_o$ are respectively inner and outer radii. $x_i$ and $x_f$ are respectively the starting and ending longitudinal positions.  Most of the coils are connected in series, and typically receive a current $I_{\textrm{main}}=32.5\,\textrm{A}$. Coils n°3 and n°5 control the field gradient at the SF1 position.}
  \label{tab:coils}
\end{table}

\newpage
\printbibliography

\end{document}
